\documentclass[preprint,epsfx,floatfix]{revtex4-1}
\usepackage{graphicx}
\usepackage{enumerate}
\usepackage{float}
\begin{document}
\title{Dynamically Varying Networks}
\title{Dynamics of Networks with Stochastically Varying Links} 
\title{Spatiotemporal Regularity in Networks with Stochastically Varying Links} 
\author{Ankit Kumar, Vidit Agrawal and Sudeshna Sinha}
\affiliation{Indian Institute of Science Education and Research
  (IISER) Mohali, Knowledge City, SAS Nagar, Sector 81, Manauli PO 140
  306, Punjab, India}

\begin{abstract}
  In this work we investigate time varying networks with complex
  dynamics at the nodes. We consider two scenarios of network change
  in an interval of time: first, we have the case where each link can
  change with probability $p_t$, i.e. the network changes occur
  locally and independently at each node. Secondly we consider the
  case where the entire connectivity matrix changes with probability
  $p_t$, i.e. the change is global. We show that network changes,
  occuring both locally and globally, yield an enhanced range of
  synchronization.  When the connections are changed slowly
  (i.e. $p_t$ is low) the nodes display nearly synchronized intervals
  interrupted by intermittent unsynchronized chaotic bursts. However
  when the connections are switched quickly (i.e. $p_t$ is large), the
  intermittent behavior quickly settles down to a steady synchronized
  state. Furthermore we find that the mean time taken to reach
  synchronization from generic random initial states is significantly
  reduced when the underlying links change more rapidly. We also
  analyse the probabilistic dynamics of the system with changing
  connectivity and the stable synchronized range thus obtained is in
  broad agreement with those observed numerically. 

\bigskip
\bigskip
\bigskip
\bigskip

\end{abstract}

\maketitle

Complex networks have gained much attention in the scientific
community over the last decade. Their importance arises from the wide
range of interesting phenomena they yield, as well as their success in
modelling systems prevalent in nature. Many mathematical models have
been proposed which capture the essential features of real networks
and generate artificial complex networks amenable to theorectical
analysis, like the small world model \cite{Watts1998} and the
scale-free networks model \cite{Barabsi15101999}. Despite their great
success, these mathematical models usually do not take into account
the dynamical nature of connections in complex systems. Since
connections in commonly found real networks often change over time
\cite{dynamic1,dynamic2,dynamic3,zanette,belykh,amritkar,sirs_vivek,explosive},
in this work we will propose a model network with time-varying links
and focus on the important question of the spatiotemporal implications
such switching connections.

First we describe our model network. We begin with a ring of $N$
nonlinear maps whose dynamics is given by the following evolution
equations \cite{cml}:
 \begin{equation}
   x^{i}_{n+1} = (1-\epsilon) f(x^{i}_n) + \frac{\epsilon}{2} (x^{l}_n  + x^{m}_n ) \label{evolution}
 \end{equation}

 Here $x_n^i$ is the state variable at site $i$ ($i=1, \dots N$) at
 discrete time $n$, and when $l=i+1$, $m=i-1$ we have nearest neighbour
 connections.  In the language of networks, index $i$ represents a node of
 degree $2$ in the network. The function $f(x)$ gives the local
 dynamics. To begin with, this is chosen to be the prototypical
 chaotic logistic map: $ x_{n+1} = rx_n(1-x_n)$, where the
 nonlinearity parameter $r$ is chosen to be $4$. The strength of
 coupling is denoted by $\epsilon$.

 We now go on to incorporate \emph{spatial} randomness in the
 connections using the algorithm proposed by Strogatz-Watt
 \cite{Watts1998}, namely we rewire a fraction $p_s$ of the regular
 nearest neighbour links on the ring to random nodes. So $l$ and $m$
 are $i+1$ and $i-1$ with probability $(1-p_s)$, and some random site
 $l,m \in (1, \dots, N)$ with probability $p_s$. So parameter $p_s$
 controls the topology of the network and it tunes the network from a
 regular ring at $p_s=0$ to a completely random network at $p_s=1$.

 Further, in this work we introduce an additional parameter $p_t$,
 which {\em reflects the dynamic nature of the connections among the
 nodes}. This parameter gives the probability of change of the links
 in the network in a given time interval, i.e. it is a measure of the
 frequency of link switches. Specifically if $p_t$ is zero, the
 connectivity matrix is static and all links are invariant in time. If
 $p_t =1$ then the connections always change in a given time
 interval. We consider the scenario where all temporal changes of the
 network keeps the spatial nature of the network invariant, namely the
 fraction of random links $p_s$ remains the same when the links
 switch.
        
 We propose two methods to change connections in a complex network:

 (i) Each site changes its connections in a time interval with
 probability $p_t$, namely, the changes occur independently and
 stochastically at the local level.

 (ii) All nodes change their connections simultaneously, i.e. the
 network as a whole changes, in a time interval with probability
 $p_t$.

 In the sections below we investigate the change in temporal behaviour
 of the dynamics at the nodes of this network, as the time-scale of
 the variation of links changes.
Through extensive numerical simulations of this dynamical network we
first obtain bifurcation diagrams with respect to coupling strength
$\epsilon$. From these bifurcation diagrams we find the critical
coupling strength $\epsilon_{sync}$ such that one obtains spatio-temporal
syncronization, namely a spatio-temporal fixed point, for $\epsilon
\ge \epsilon_{sync}$ .

\begin{figure}[ht]
   	\begin{minipage}{0.49\linewidth}
\includegraphics[width=1\textwidth,height=70mm]{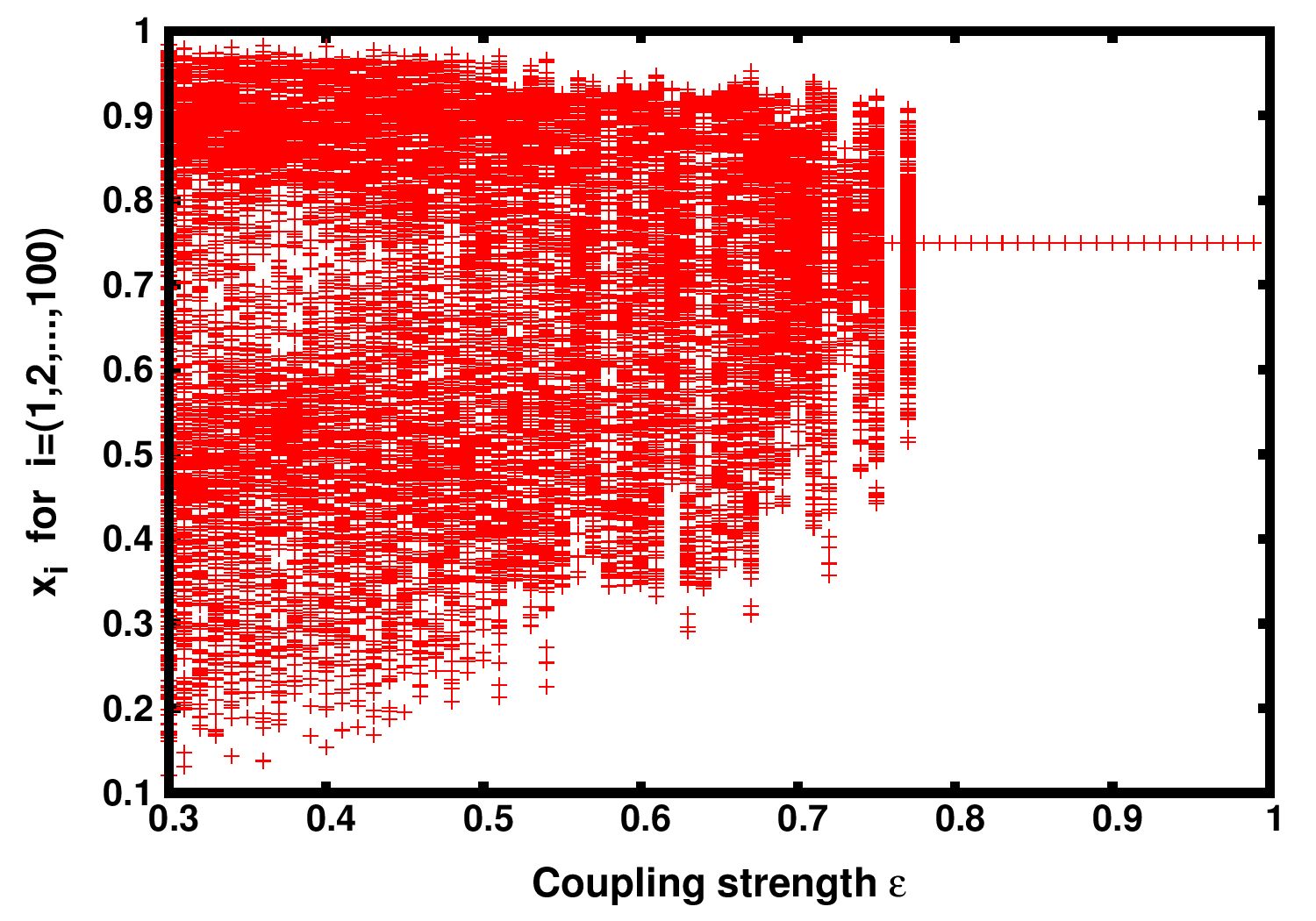}
	\end{minipage}
	\hfill
	\begin{minipage}{0.49\linewidth}
		\includegraphics[width=1\textwidth, height=70mm]{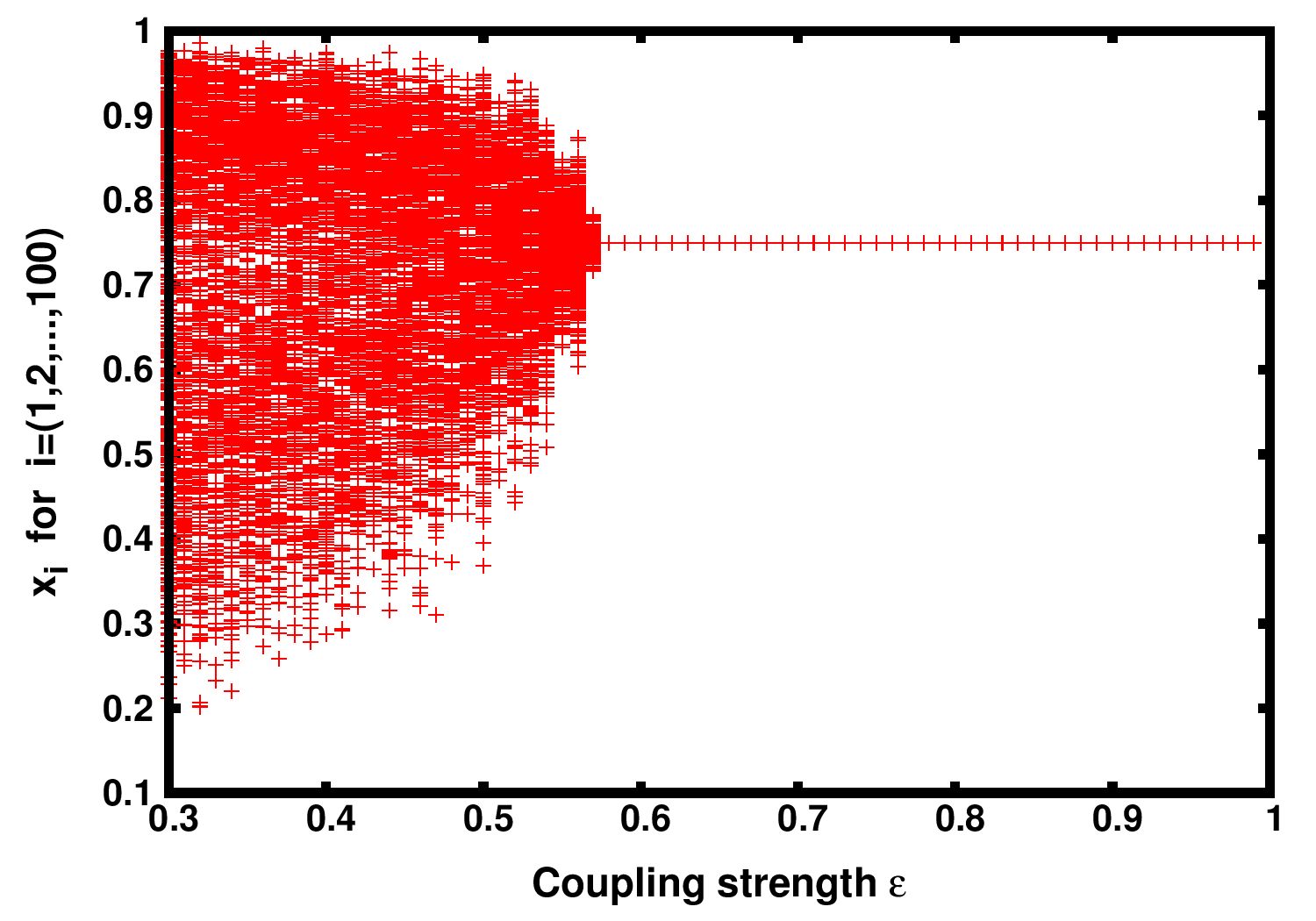}
	\end{minipage}
		\caption{ Bifurcation diagram displaying the state of
                  the network ($x_n (i), i = 1, \dots N$, with
                  $N=100$) over $5$ time steps, after transience of
                  $4000$ steps, starting from a random initial
                  condition, for the case of global link changes. Here
                  the fraction of random links is $p_s=0.80$ and the
                  link rewiring probability $p_t$ is $0.01$ (left)
                  $0.95$ (right).}
\label{bif}
   \end{figure}

It is clearly evident from Fig. \ref{bif} 
the critical coupling strength $\epsilon_{sync}$ decreases as link
switching probability $p_t$ increases.  Namely, as the {\em
  probability of changing links increases, the range for the
  spatio-temporal fixed point increases}. So a more dynamic web of
links is more favourable for inducing spatiotemporal regularity in
coupled chaotic systems.
Surprisingly we found that both methods show similar qualitative
effects on the dynamics of the nodes, despite the fact that the local
method involves incremental changes in connectivity and the
global case implies sudden and large changes in connectivity.

\bigskip

We now describe the degree of syncronization \cite{barahona, hong,
  intermediate} in the system quantitatively through the
synchronization error function defined as
\begin{equation}
\ Z(n) = \frac{1}{N}\sum_{i=1}^N[x_n(i) - x_{mean}]^2
\end{equation}
averaged over time \emph{n} and calculated after transient time, with
$x_{mean}$ being the mean value of $x(i) = 1,2, \dots N$ at a given time
step $n$.

\begin{figure}[ht]
   	\begin{minipage}{0.49\linewidth}
\includegraphics[width=1\textwidth, height=70mm]{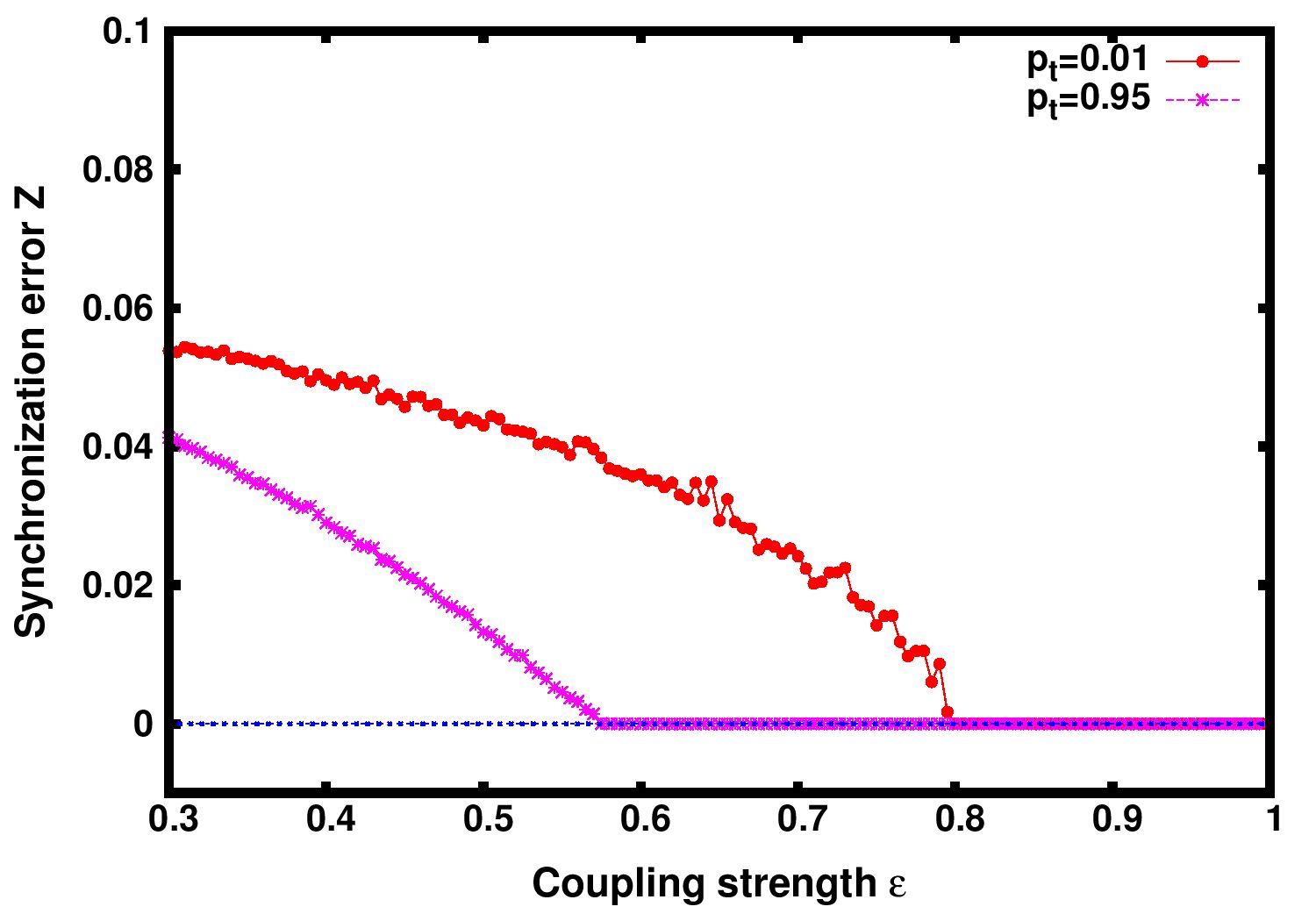}
	\end{minipage}
	\hfill
	\begin{minipage}{0.49\linewidth}
 \includegraphics[width=1\textwidth,height=70mm]{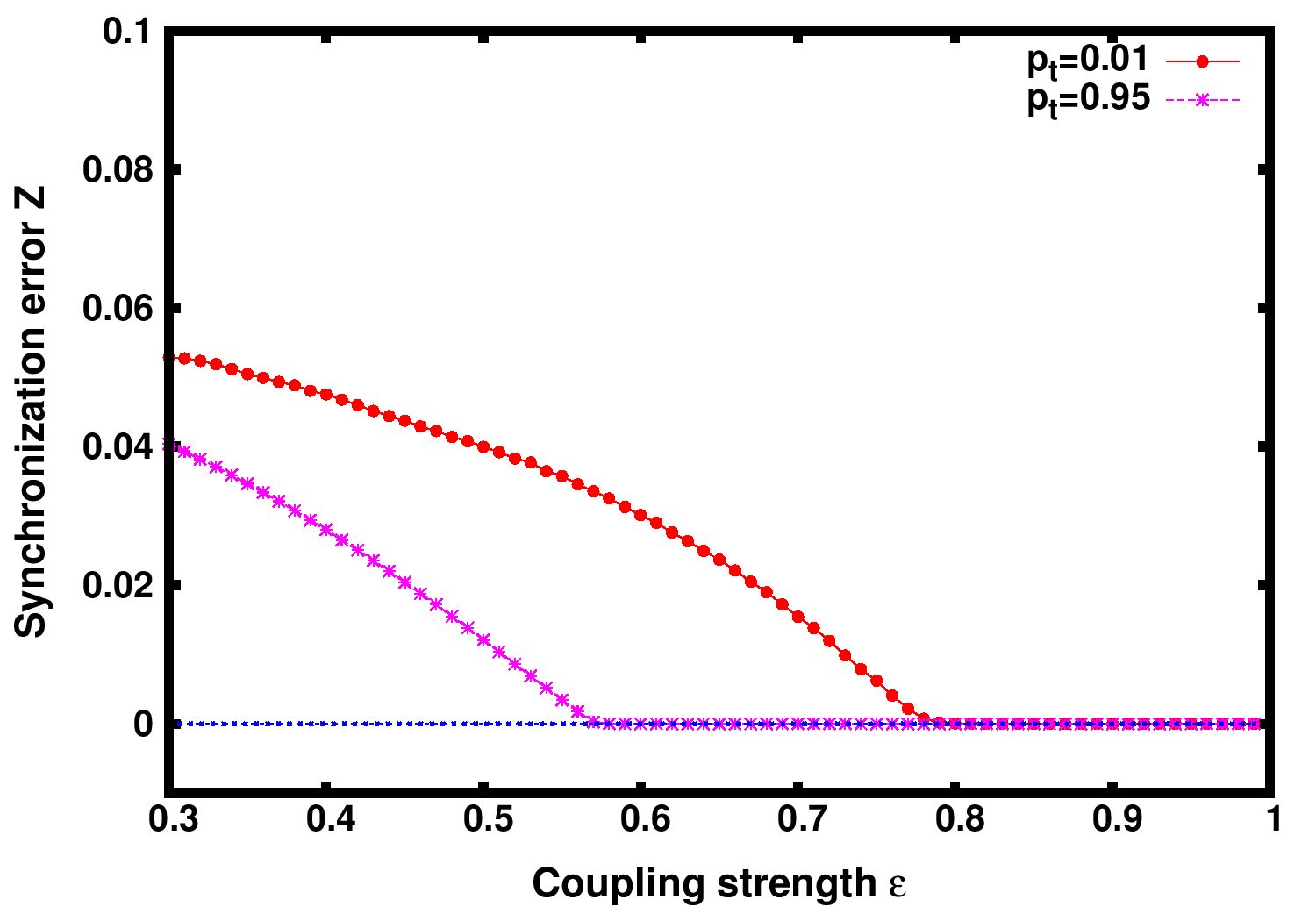}
	\end{minipage}
		\caption{
                  Synchronization error $Z$ as a function of coupling
                  strength for link switching probability $p_t=0.01$
                  (pink), $0.95$ (red), for the case of (right) global 
                  and (left) local link changes. Here fraction of
                  random links $p_s=0.80$, system size $N=100$ and $Z$
                  was obtained by averaging over $4000$ time steps and
                  $100$ different initial realizations (see text).}
\label{sync}
   \end{figure}

Fig. \ref{sync} displays the variation of synchronzation error
(averaged over space and time) with respect to the coupling strength
$\epsilon$, for the case of both local and global link rewiring. It
can be clearly seen that as the connection network becomes more
dynamic, the range of complete syncronization increases. Furthermore,
for both the connection rewiring scenarios the qualitative results are
similar.

\bigskip

\begin{figure}[ht]
\includegraphics[width=0.75\textwidth,height=70mm]{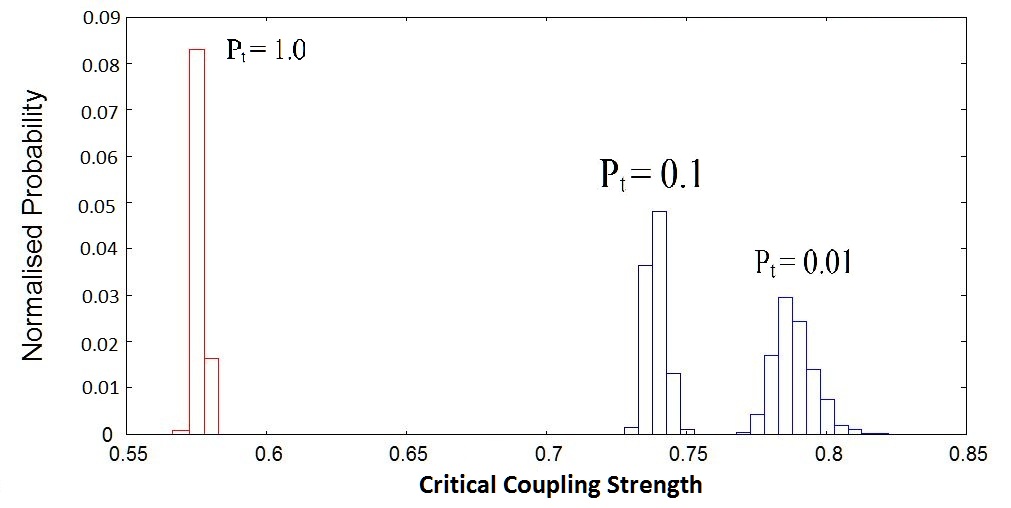}
		\caption{
                  Distribution of critical coupling strengths
                  $\epsilon_{sync}$ obtained from sampling $100$
                  different random initial realizations.  Here
                  fraction of random links $p_s=0.80$, network size
                  $N=100$ and link switching probability $p_t = 0.01$,
                  $0.1$ and $1.0$.}
		\label{subfig: histo01} 
   \end{figure}

Now, we consider the scenario where link changes are infrequent,
namely a network near the static limit, with $p_t$ close to zero. Here
one obtains a range of critical coupling strengths, with
$\epsilon_{sync}$ being strongly dependent on the initial confguration
of links. A deeper understanding is gained by studying the
distribution of $\epsilon_{sync}$, at fixed $p_t$ and $p_s$, for
different initial realizations, as displayed for representative cases
in Fig. \ref{subfig: histo01}. It is clearly seen from the numerics
that when we are closer to the static limit there is a spread in
values of $\epsilon_{sync}$. As the system becomes more dynamics,
i.e. as $p_t$ increases, we observe that the spread of
$\epsilon_{sync}$ narrows considerably, converging rapidly to the
average value $\langle \epsilon_{sync} \rangle$. This is a reflection
of the more effective ``self-averaging'' arising from dynamically
changing network configurations as the system evolves for larger
$p_t$.

Further, the average critical coupling strength $\langle
\epsilon_{sync} \rangle$ also shifts to a smaller value with
increasing link switching probability $p_t$. This implies that lower
coupling strengths are necessary to bring about synchronization when
the link changes are more rapid.

\begin{figure}[ht]
\includegraphics[width=0.75\textwidth, height=70mm]{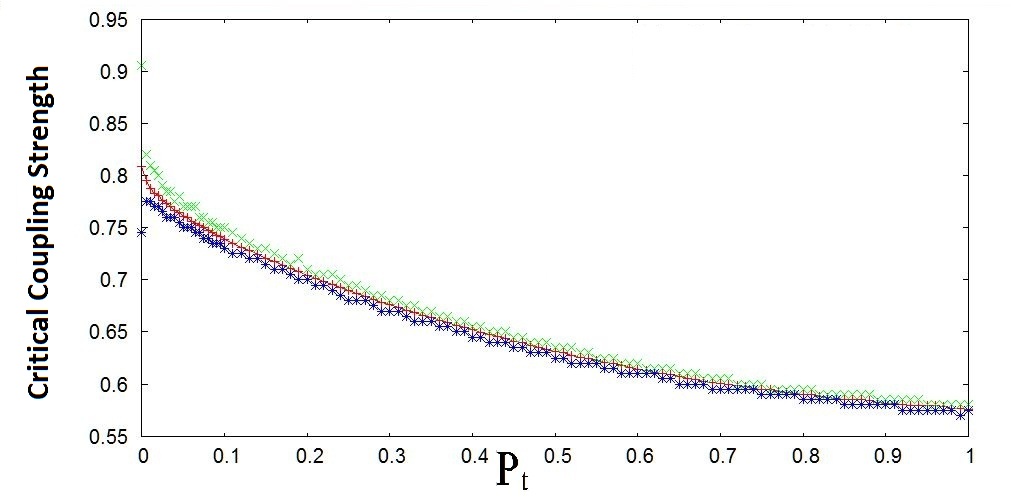}
		\caption{
                  Variation of the average critical coupling strength
                  $\langle \epsilon_{sync} \rangle$ (red), maximum
                  critical coupling strength $\epsilon_{max}$ (green)
                  and minimum critical coupling strength
                  $\epsilon_{min}$ (blue), obtained from sampling
                  $100$ different random initial realizations, with respect
                  to link switching probability $p_t$. Here fraction
                  of random links $p_s=0.80$ and network size $N=100$.}
		\label{subfig: spread01}
   \end{figure}

Next we show the variation of the average critical coupling strength
$\langle \epsilon_{sync} \rangle$ with respect to the probability of
link change $p_t$. Figs. \ref{subfig: spread01} display the average
critical coupling strength, the maximum value of critical coupling
strength $\epsilon_{max}$ and the mimimum value of critical coupling
strength $\epsilon_{min}$. It is evident from the plot that $\langle
\epsilon_{sync} \rangle$ diplays a clear trend under increasing $p_t$, even
at low $p_t$ when there is considerable separation in the
$\epsilon_{min}$ and $\epsilon_{max}$ values.
Results from the local rewiring scheme is diplayed here. Similar
phenomena is observed for the case of global link changes as well.

\bigskip

{\em Intermittent approach to synchronization:}\\
 
\begin{figure}[ht]
   	\begin{minipage}{0.49\linewidth}
		\includegraphics[width=1.0\textwidth, height=70mm]{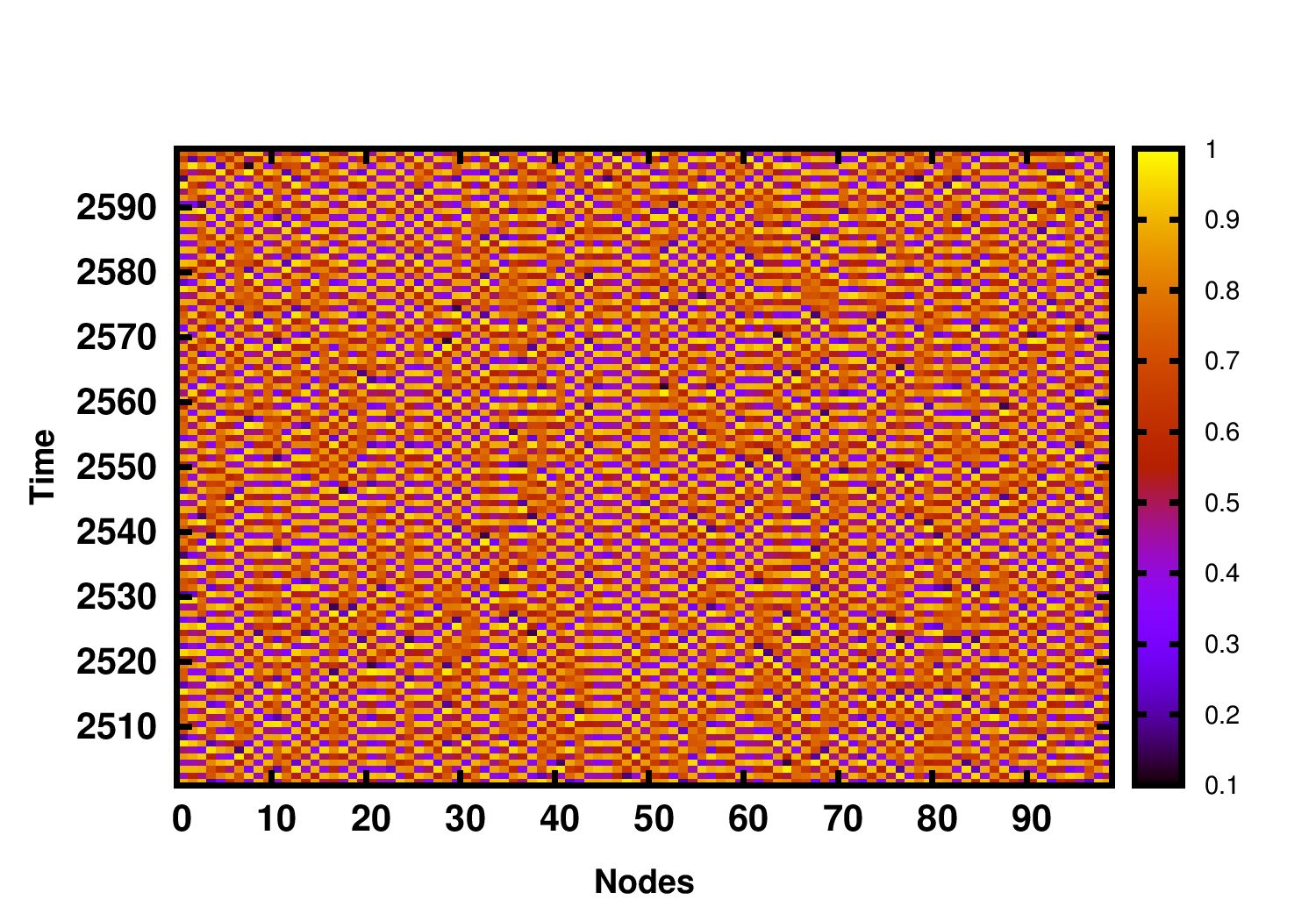}
	\end{minipage}
	\hfill
  	\begin{minipage}{0.49\linewidth}
   		\includegraphics[width=1.0\textwidth,height=70mm]{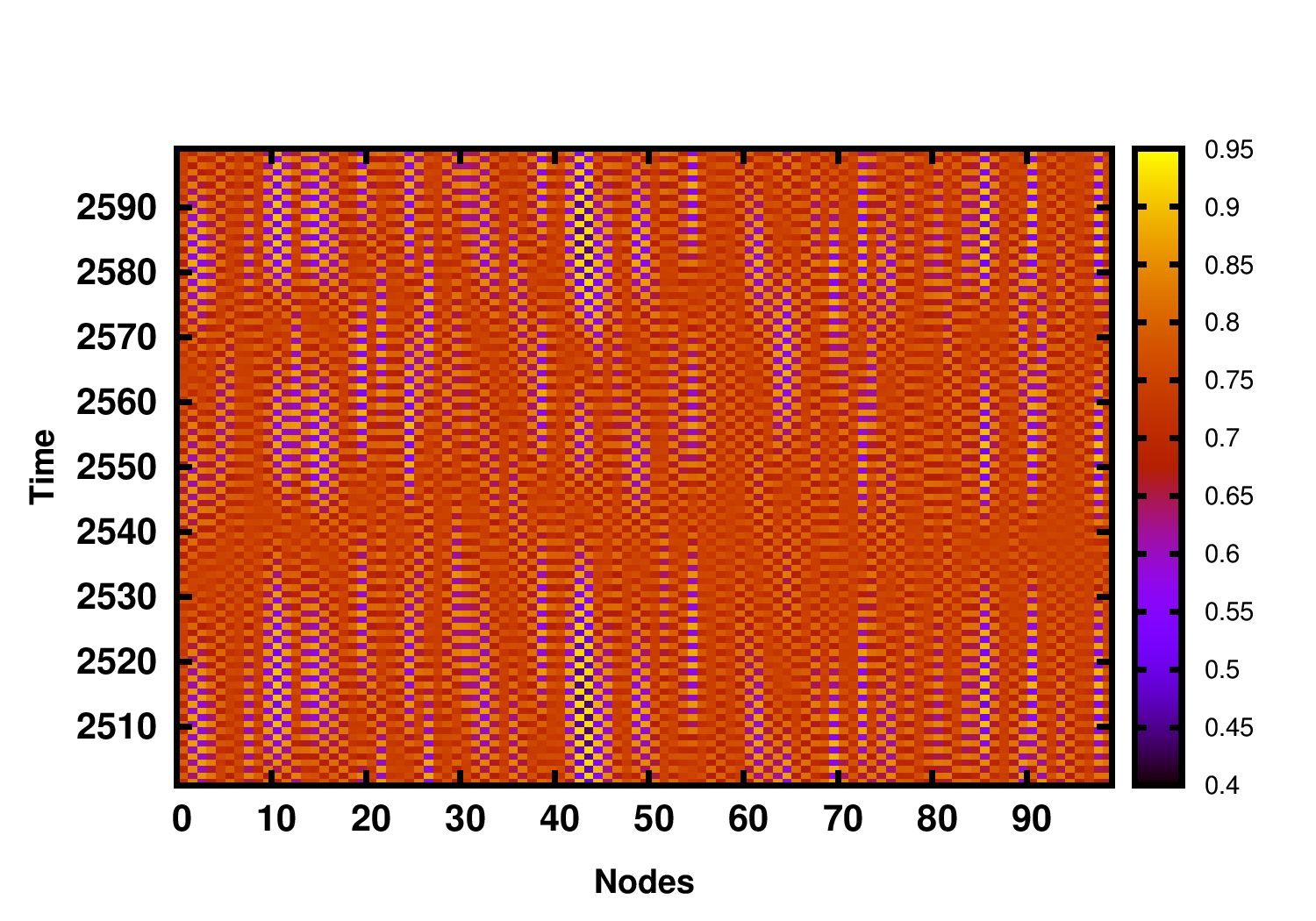}
	\end{minipage}
		\caption{
                  Spatiotemporal evolution of the system of size
                  $N=100$, with fraction of random links $p_s=0.80$
                  and coupling strength $\epsilon$ equal to $0.24$
                  (left) and $0.75$ (right). Here links are changed
                  globally with probability $p_t=0.01$.}
		\label{density} 
   \end{figure}

  Examining the spatiotemporal evolution of the network, as displayed
  in Fig. \ref{density} reveals the following feature: one can see
  low coupling strengths the system exhibits spatio-temporal chaos
  (cf. left panel of Fig. \ref{density}).  However as coupling
  strength increases
  one observes a dynamical regime in which the system displays
  intermittent behaviour. Fig. \ref{density} (right) shows the
  spatio-temporal evolution of one such representative regime,
  exhibiting syncronized periods with burst of unsyncronized
  behaviour.  Similar qualitative dynamical patterns are obtained for
  both local and global network changes.

Now to study these intermittent patterns in greater detail, we define
a parameter $L_{intermittent}$ which is the average length of
intermitent behaviour in time. Quantitatively this length represents
the time between the the first event of near complete spatiotemporal
syncronization and the last observed unsynchronized
burst. Specifically, without loss of generality, we consider a system
synchronized if the synchronization error $Z < 10^{-5}$.

\begin{figure}[ht]
   		\includegraphics[width=0.75\textwidth,height=55mm]{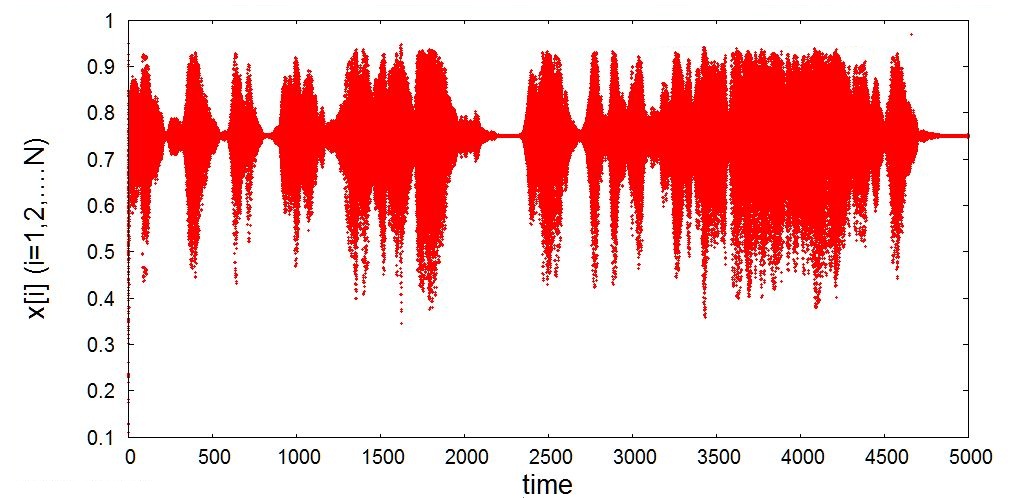}
		\caption{
                  Time evolution of the system ($x_n (i), i = 1, \dots
                  N$) over $5000$ time steps after transience,
                  starting from one random initial realization, for
                  the case of local link change. Here the fraction of
                  random links $p_s=0.80$, link switching probability
                  $p_t= 0.01$, coupling strength $\epsilon = 0.76$ and
                  network size $N=100$. Qualitatively similar
                  intermittenct approach to synchronization is
                  observed for the case of global connection changes
                  as well.}
		\label{subfig: int02} 
   \end{figure}

\begin{figure}[ht]
\includegraphics[width=0.45\textwidth, height=45mm]{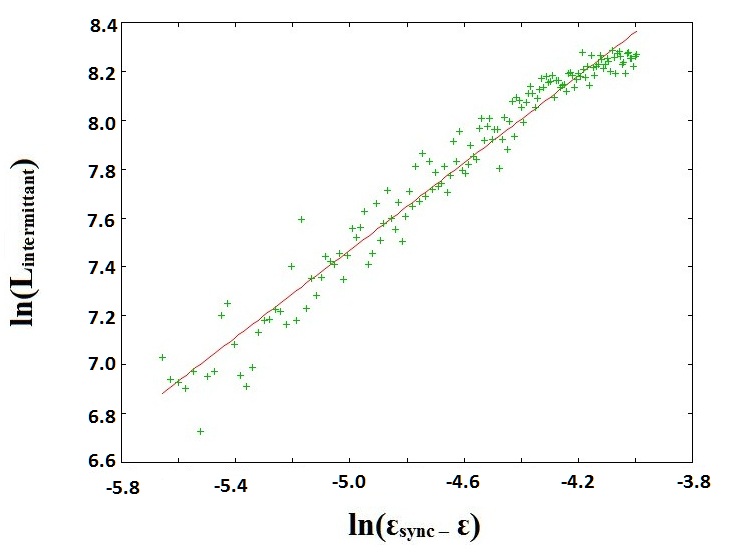}
\includegraphics[width=0.45\textwidth, height=45mm]{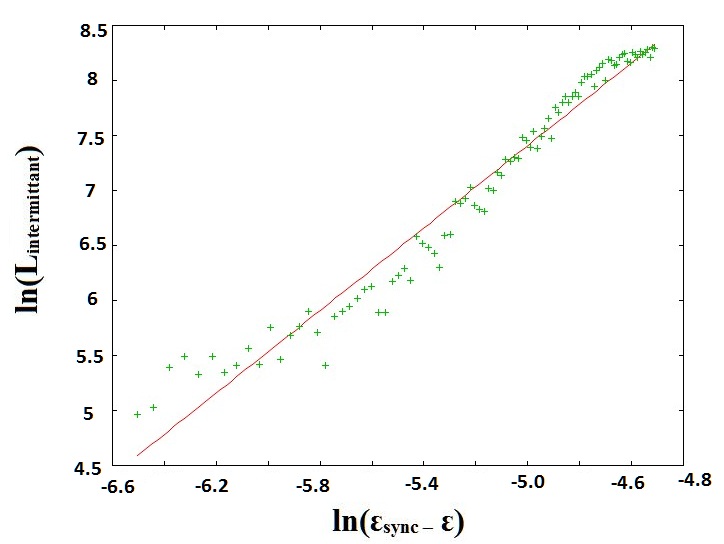}
\caption{
  Average length of intermittency $L_{intermitent}$ vs. distance of
  coupling strength from critical strength $(\epsilon_{sync} -
  \epsilon)$, for the case of local link chnages, with fraction of
  random links $p_s=0.8$, link switching probability $p_t=0.01$ (left)
  and $p_t=0.1$ (right).  Here network size $N=100$, and
  $L_{intermittent}$ is obtained by averaging over $100$
  realizations. Power law fit of $L_{intermittent}$ to
  $(\epsilon_{sync} - \epsilon)^{\mu}$ (solid line), with $\mu =
  1.8759 \pm 0.0385$ for $p_t = 0.1$ and $\mu = 0.8917 \pm 0.0146$ for
  $p_t =0.01$, is also displayed.}
		\label{subfig: int01}
\end{figure}

In Fig. \ref{subfig: int02} we present a simple illustration of one
such case. Here the first syncronized stretch is seen at $t \approx
800$. Subsequently one obtains desyncronized bursts, followed by
synchronized intervals, over a period of time. Finally at $t \approx
4600$ the last burst of unsyncronized behaviour is seen after which
the system remains syncronized up to the limits of the simulation time
($t \approx 10^4$). For the completely chaotic region at low coupling
strengths, and the completely syncronized region at high coupling
strengths there is no intermittency as evident from $L_{intermittent}
\rightarrow 0$. However, for a range of coupling approaching the
critical coupling strength, the average length of the intermittent
period increases as a power law with respect to distance from the critical
point (see Fig. \ref{subfig: int01} for representative examples). It
is further evident that increasing the probability of changing links
reduces the range of coupling strengths over which this intermittent
approach to synchronization is observed.


\bigskip

{\em Mean time to reach the synchronized state:}\\

\begin{figure}[ht]
\includegraphics[width=0.7\textwidth]{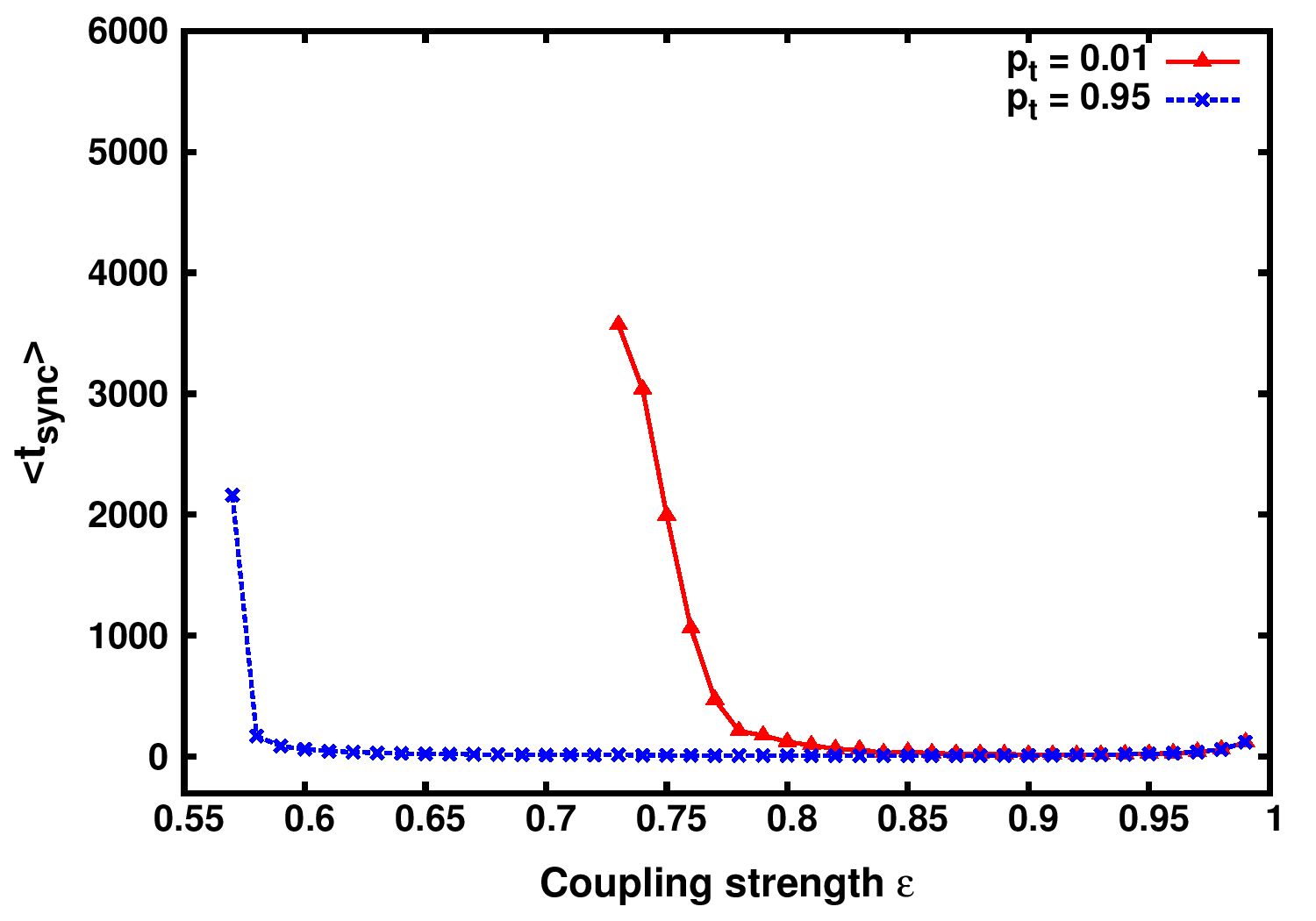}

\caption{
  Mean synchronization time $\langle t_{sync} \rangle$ for the system
  of size $N=100$, with fraction of random links $p_s=0.80$ and link
  change probability $p_t = 0.01$ (left) and $p_t = 0.95$
  (right). Here links are changed globally with probability $p_t$.}
			\label{mean_sync} 
	   \end{figure}

We have also investigated the average time taken by the system,
starting from generic random initial conditions, to reach the
synchronized state. The results are displayed for two representative
cases in Fig. \ref{mean_sync}. It is clear that more rapid link changes
lead to much shorter transience. So the system very quickly reaches
the spatiotemporal fixed point when the connections are varying fast.

\bigskip

{\em Analysis:}\\

We now analyse the system to account for the much enhanced stability
of the spatiotemporal fixed point under rapidly changing
connections. The only possible solution for a spatiotemporally
synchronized state here is $x_n (i) = x^*$, where $x^* = f(x^*)$ is
the fixed point solution of the local map. For the case of the
logistic map this is $x^* = 4 x^* (1-x^*) = 3/4$.

To calculate the stability of the network with all sites at $x^*$, we
construct an {\it probabilistic evolution rule} for the state of the
nodes.  In this mean field-like version of the dynamics, the effective
influence of the random connections on the local dynamics is given by
$p_{eff}$, and the influence of the nearest neighbours is given by
$(1-p_{eff})$, where $p_{eff}$ is determined by the link change
probability $p_t$ as well as the fraction of random sites $p_s$.

In terms of $p_{eff}$ the averaged evolution equation of node $j$ ($j
= 1, \dots N$) is
\begin{equation}
x_{n+1} (j) =
(1-\epsilon) f (x_n(j)) + (1-p_{eff})\frac{\epsilon}{2} ( x_{n}(j+1)+x_n (j-1) )+
p_{eff} \frac{\epsilon}{2} (x_n (\zeta) + x_n(\eta))
\label{peff}
\end{equation}
Where $\zeta $ and $\eta $ are two random sites ($\zeta, \eta \in [1,
N]$).

Now in order to calculate the stability of the synchronized state, we
linearize Eq.~\ref{peff}, by considering $x_n(j) = x^* + h_n(j)$, and
expanding to first order:
\begin{equation}
h_{n+1}(j) = (1 - \epsilon)f'(x^*)h_{n}(j) + (1-p_{eff}) \frac{\epsilon}{2} \left\{
h_n (j+1) + h_n (j-1) \right\} + p_{eff} \frac{\epsilon}{2} \left\{ h_n (\zeta) + h_n (\eta) \right\}
\end{equation}
Considering the sum over uncorrelated random neighbours to be equal to
zero, one obtains the approximate evolution equation:
\begin{equation}
h_{n+1}(j) = (1 - \epsilon)f'(x^*)h_{n}(j) + (1-p_{eff}) \frac{\epsilon}{2} \left\{
h_n (j+1) + h_n (j-1) \right\}.
\end{equation}

For stability considerations one can diagonalize the above expression
using a Fourier transform $h_n(j) = \sum_{q} \phi_n (q) \exp(i \ j q)$,
where $q$ is the wavenumber and $j$ is the site index, which yields
the following growth equation:

\begin{equation}
\frac{\phi_{n+1} (q)}{\phi_n (q)} =  f'(x^*)(1 - \epsilon) + \epsilon (1-p_{eff}) \cos q
\end{equation}
with $q$ going from $0$ to $\pi$. Specifically, for the case of the
chaotic logistic map at $r=4$ we have $f^{\prime} (x^*) = -2$. So the
magnitude of the growth coefficient that appears in the above
expression is smaller than $1$q, if and only if
\begin{equation}
\frac{1}{1+p_{eff}} < \epsilon < 1
\end{equation} 

This inequality then yields the coupling strength $\epsilon_{sync}$
after which the spatiotemporal fixed point gains stability to be:
\begin{equation}
\epsilon_{sync} = \frac{1}{1+p_{eff}}
\end{equation}

Further the range of the spatiotemporal fixed point $\cal{R}$ is given by:

\begin{equation}
{\cal R} = 1-\epsilon_{sync} = \frac{p_{eff}}{1+p_{eff}}
\end{equation}

Now $p_{eff}$ is the probability that the links are different from
time to time. So $p_{eff}$ must be directly proportional to the
probability of random rewiring $p_t$ and the fraction of random links
$p_s$. Starting with the ansatz that $p_{eff} = f (p_s \ p_t)$, where
function $f$ is a power law, gives:
\begin{equation}
{\cal R} \sim \frac{(p_s \ p_t)^{\nu}}{1+(p_s \ p_t)^{\nu}} 
\label{ansatz}
\end{equation}

\begin{figure}[ht]
\includegraphics[width=1\textwidth]{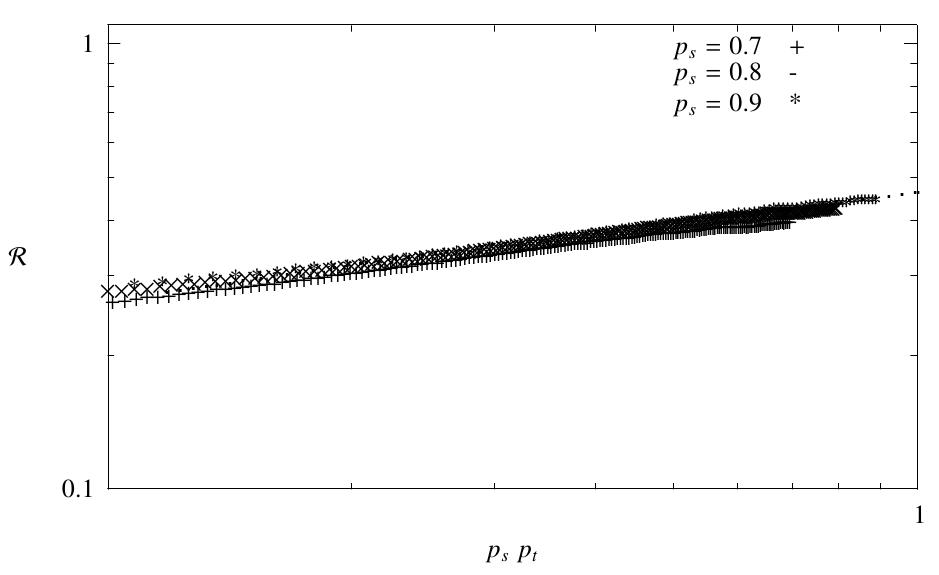}
\caption{Variation of range of the spatiotemporal fixed point
  $\cal{R}$ with respect to the product of fraction of random links
  and probability of link change $p_s p_t$, for the case of local link
  changes. Fitting this to Eq.~\ref{ansatz} in the range $p_s p_t \in
  [0.1:1]$ yields $\nu \sim 0.4$ (shown by dotted line).}
		\label{scale}
	\end{figure}

Fig.~\ref{scale} displays the dependence of the range of the
spatiotemporal fixed point obtained numerically for the case of local
link change, on $p_s p_t$. Fitting this to Eq.~\ref{ansatz} yields
$\nu \sim 0.4$ for the range $0.1 < p_s p_t < 1$. The range of the
spatiotemporal fixed point for the case of global changes can also be
fit to the same functional form, with best fit yielding $\nu \sim 0.3$
in a similar range of $p_s p_t$.

\bigskip

{\em Generality of the Results:}\\

In order to gauge the generality of our results, we also analysed a
network of Exponential Maps (also known as the Ricker Map). These are
given by the dynamical equation:
\begin{equation}
f(x) = x \ e^{r (1-x)}
\end{equation}
In the results presented here we take the nonlinearity parameter $r$ to be $2.6$, where the map is strongly chaotic.

\begin{figure}[ht]
		\includegraphics[width=0.75\textwidth, height=70mm]{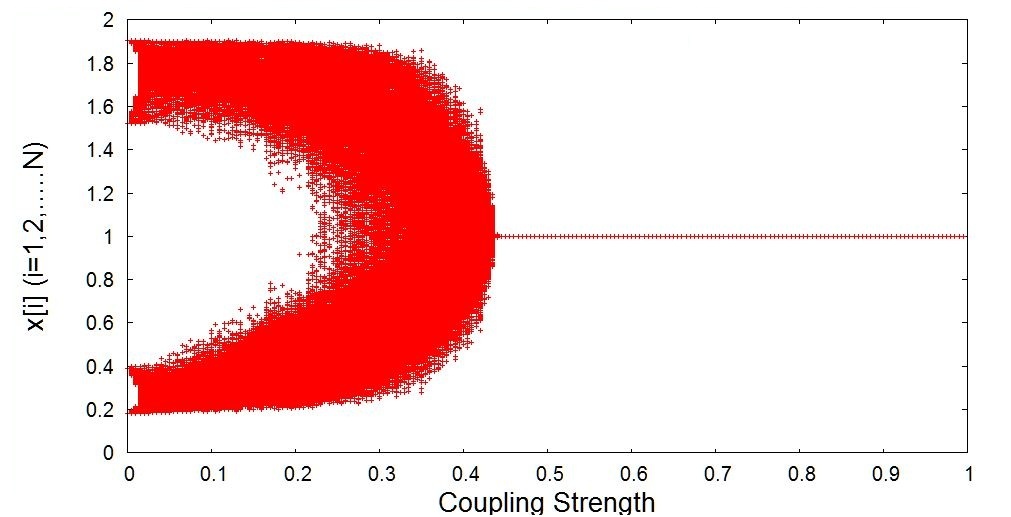}
		\caption{
                  Bifurcation diagram displaying the state of the
                  network of Ricker (Exponential) maps ($x_n (i), i =
                  1, \dots N$, with $N=100$) over $5$ time steps,
                  after transience of $4000$ steps, starting from a
                  random initial condition, for the case with local
                  link changes. Here the fraction of random links is
                  $p_s=0.80$ and the link rewiring probability is
                  $p_t=0.95$.}
		\label{subfig: bif03}
	\end{figure}

Representative results are shown in Fig. \ref{subfig: bif03}.
Clearly, spatiotemporal synchronization is obtained at coupling
strengths $\epsilon > \epsilon_{sync}$, where $\epsilon_{sync} = 0.48$
for $p_t = 1$.

\begin{figure}[ht]
   		\includegraphics[width=0.75\textwidth,height=70mm]{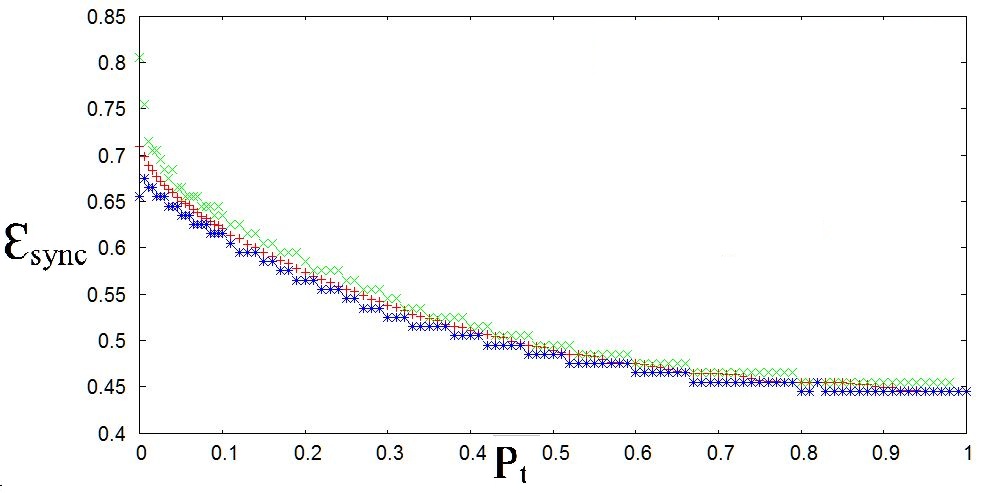}
		\caption{
                  Critical coupling strengths $\epsilon_{sync}$ vs
                  link switching probability $p_t$, for fraction of
                  random links $p_s = 0.80$ for a network of Ricker
                  (Exponential) maps.}
		\label{subfig: spread02}
   \end{figure}

Further we calculate the variation of $\langle \epsilon_{sync}
\rangle$, maximum $\epsilon_{sync}$ and minimum $\epsilon_{sync}$ with
respect to link rewiring probability $p_t$, and the results are
displayed in Fig. \ref{subfig: spread02}. It is evident that the
qualitative picture that emerges is the same as in a network of
chaotic logistic maps. Namely, we again observe a wider separation
between $\epsilon_{max}$ and $\epsilon_{min}$ at very low $p_t$
(i.e. close to the static limit), and this shrinks rapidly as $p_t$
increases. As before, we also find a smooth decreasing trend for
$\langle \epsilon_{sync} \rangle$ with increasing $p_t$. So it is clear
that more frequent link changes enhances the range of spatiotemporal
synchronization, and the critical coupling strength necessary to
obtain the spatiotemporal fixed point is lower in networks with
faster variation in connectivity.

   In summary, in this work we have investigated time varying networks
   with complex dynamics at the nodes. We considered two scenarios of
   network change in an interval of time: first, we have the case
   where each link can change with probability $p_t$, i.e. the network
   changes occur locally and independently at each node. Secondly we
   considered the case where the entire connectivity matrix changes
   with probability $p_t$, i.e. the change is global. We demonstrated
   that network changes, occuring both locally and globally, yield an
   enhanced range of synchronization.  When the connections are
   changed slowly (i.e. $p_t$ is low) the nodes display nearly
   synchronized intervals interrupted by intermittent unsynchronized
   chaotic bursts. However when the connections are switched quickly
   (i.e. $p_t$ is large), the intermittent behavior quickly settled
   down to a steady synchronized state. Furthermore we found that the
   range of synchronization increases significantly with the
   probability of network change $p_t$. Additionally the system
   reaches the synchronized state much more rapidly for the case of
   switched links. Thus our results highlight the strong effects of
   time-varying connections on the nodal dynamics, and our principal
   observations are relevant to the understanding of temporal networks
   in general.

\vskip 1.5in


\end{document}